# Enhancement of microcavity lifetimes using highly dispersive materials

Marin Soljačić[a], Elefterios Lidorikis[a], Lene Vestergaard Hau[b], J.D. Joannopoulos[a]

(a) Physics Department, MIT, Cambridge, MA 02139
(b) Physics Department, Harvard University, Cambridge, MA 02138

We show analytically, and numerically that highly-dispersive media can be used to drastically increase lifetimes of high-Q microresonators. In such a resonator, lifetime is limited either by undesired coupling to radiation, or by intrinsic absorption of the constituent materials. The presence of dispersion weakens coupling to the undesired radiation modes and also effectively reduces the material absorption.

Microcavities with long life-times, and very small modal volumes (i.e. very narrow transmission resonance widths $\Gamma_{TRANS}$) have important applications in many different fields including: photonics [1,2], non-linear optics [3], biosensing [4], cavity quantum electrodynamics [5,6], and novel frequency standards [7]. For many applications, on-chip-resonators are highly preferred [8]. Limits to increasing lifetimes of such micro-resonators are most often determined by their intrinsic losses: absorption of the constituent material (which determines the absorption quality factor $Q_{ABS} \equiv \omega_{RES}/(2\Gamma_{ABS})$), or undesired coupling to radiation modes due to imperfect confinement (which determines the radiation $Q_{RAD}$). The quality factor of a resonator $Q_{TRANS} = 1 \Big/ \left( \dfrac{1}{Q_{RAD}} + \dfrac{1}{Q_{ABS}} + \dfrac{1}{Q_{IO}} \right)$ cannot be larger than the smaller of $Q_{RAD}$ and $Q_{ABS}$ ($Q_{IO}$ is determined by the coupling to the input and output). On a separate front, recent work [9,10,11] has shown how the transmission curve of a perfect macroscopic ring cavity, coupled only to its input and output ports can be significantly narrowed by insertion of highly dispersive media (in that case, using electromagnetically induced transparency (EIT)). It is tempting to speculate whether using highly-dispersive media could be used to significantly narrow transmission resonance widths of high-Q microresonators. Naively, one might think that this approach cannot work: if dispersion increases the lifetime of a resonator, that means that the light residing in the resonator has more time to interact with the absorptive material, and more time to couple to the undersired radiation modes. Therefore, it seems that one would again be limited by $Q_{ABS}$, and $Q_{RAD}$, the same as before. In this manuscript, we show, that this picture is not correct: inserting highly dispersive material into a cavity drastically increases all of $Q_{RAD}$, $Q_{ABS}$, and $Q_{IO}$, and this phenomenon could therefore be used to design micro-resonators with lifetimes orders of magnitude larger than what was previously possible.

For definiteness, imagine a resonator, with one input, and one output channel, with equal input and output couplings. The material from which the resonator is made is approximately non-dispersive, but it has some finite absorption. Imagine further that the confinement of the resonator is not perfect, so the resonator is also coupled to undesired radiation modes. The transmission of this resonator can be modeled [12] as:

$$\frac{P_{OUT}(\omega)}{P_{IN}(\omega)} = \frac{\Gamma_{IO}^2}{(\omega - \omega_{RES})^2 + (\Gamma_{RAD} + \Gamma_{ABS} + \Gamma_{IO})^2} \quad , \qquad (1)$$

where $P_{OUT}$, and $P_{IN}$ are outgoing and incoming powers, $\omega_{RES}$ is the resonant frequency, $\Gamma_{ABS}$ is the absorption-decay width, $\Gamma_{RAD}$ is the radiation-decay width, and $\Gamma_{IO}$ is the width due to the coupling with input and output. As long as $\Gamma_{ABS}, \Gamma_{RAD} \ll \Gamma_{IO}$, the transmission width of this system can efficiently be lowered by decreasing the coupling to input and output ($\Gamma_{IO}$). However, because of non-zero $\Gamma_{ABS}$ and $\Gamma_{RAD}$, this program cannot be followed indefinitely: the ultimate limit to the transmission $Q_{TRANS} \equiv \omega_{RES}/2(\Gamma_{RAD} + \Gamma_{ABS} + \Gamma_{IO})$ of this system is set by: $Q_{TRANS} < \omega_{RES}/2(\Gamma_{RAD} + \Gamma_{ABS})$. Note also that as we approach the limiting $Q_{TRANS}$, according to Eq.(1), the peak transmission drops rapidly to zero.

For pedagogical reasons, consider first changing the index of refraction inside such a cavity by a small $\delta n$. According to perturbation theory, the only effect of this $\delta n$ will be to change the resonance frequency $\omega_{RES}$, thereby sliding the whole transmission curve in Eq.(1) sideways: $\omega_{RES} \rightarrow \omega_{RES}[1-\delta n \sigma/n(\omega_{RES})]$, where $\sigma$ is the fraction of the **D**-energy of the cavity mode contained in the region where $\delta n$ is applied:

$$\sigma \equiv \left[ \int_{VOL\_\delta n} d^3x \varepsilon(r)|E(r)|^2 \right] \Big/ \left[ \int_{VOL\_MODE} d^3x \varepsilon(r)|E(r)|^2 \right].$$

Next, consider replacing the material from which the cavity is made, with a material that has the same $n(\omega_{RES})$, but is now highly dispersive; we can use the same perturbation theory to determine what happens. However, in this case, every frequency $\omega$ experiences a different shift of resonance frequency: that is, $\omega$ perceives being in a system in which the induced $\delta n$ is given by:

$$\delta n(\omega) = n(\omega) - n(\omega_{RES}) \approx (\omega - \omega_{RES}) \frac{dn}{d\omega}\bigg|_{\omega_{RES}} ; \tag{2}$$

note that clearly $\delta n(\omega_{RES})=0$. This $\delta n$ is small, since we are interested only in behavior of frequencies close to $\omega_{RES}$; so, we are justified in using a perturbative approach for studying this problem. It is convenient to express everything in terms of the group velocity $v_G(\omega_{RES}) = c \Big/ \left( n(\omega_{RES}) + \omega_{RES} \frac{dn}{d\omega}\bigg|_{\omega_{RES}} \right)$ [13] of the dispersive material, and model the $v_G$ as being nearly constant over the narrow spectral band-width of the cavity [14]: $\frac{dn}{d\omega}\bigg|_{\omega_{RES}} \approx \left( \frac{c}{v_G} - n(\omega_{RES}) \right) \Big/ \omega_{RES}$. Finally, according to our perturbation theory, each frequency $\omega$ perceives a different effective $\tilde{\omega}_{RES}(\omega) \equiv \omega_{RES} - \frac{\sigma(\omega - \omega_{RES})}{n(\omega_{RES})} \left( \frac{c}{v_G} - n(\omega_{RES}) \right)$.

Plugging $\tilde{\omega}_{RES}$ in place of $\omega_{RES}$ in Eq.(1), and slightly re-arranging we obtain:

$$\frac{P_{OUT}(\omega)}{P_{IN}(\omega)} = \frac{\Gamma_{IO}^2 \Big/ \left[1 + \frac{\sigma}{n(\omega_{RES})}\left(\frac{c}{v_G} - n(\omega_{RES})\right)\right]^2}{(\omega - \omega_{RES})^2 + (\Gamma_{RAD} + \Gamma_{ABS} + \Gamma_{IO})^2 \Big/ \left[1 + \frac{\sigma}{n(\omega_{RES})}\left(\frac{c}{v_G} - n(\omega_{RES})\right)\right]^2}. \tag{3}$$

According to Eq.(3), every single decay mechanism out of this resonator ($\Gamma_{RAD}$, $\Gamma_{ABS}$, and $\Gamma_{IO}$) gets suppressed by the *same* factor: $\left[1 + \frac{\sigma}{n(\omega_{RES})}\left(\frac{c}{v_G} - n(\omega_{RES})\right)\right]$, and thereby $Q_{TRANS}$ is increased by the same factor:

$$Q_{TRANS} \rightarrow Q_{TRANS}\left[1 + \frac{\sigma}{n(\omega_{RES})}\left(\frac{c}{v_G} - n(\omega_{RES})\right)\right], \qquad (4)$$

while $T_{PEAK}$ remains unchanged.

This increase of the quality factor can be physically understood as follows. According to Eq.(2), $\omega > \omega_{RES}$ experiences $\delta n > 0$, thereby perceiving a resonance curve shifted to the left (meaning lower transmission than otherwise). Similarly, $\omega < \omega_{RES}$ experiences $\delta n < 0$, again implying a reduction of transmission since $\omega < \omega_{RES}$ perceives the resonance curve as being shifted to the right. Therefore, the final perceived transmission width is severely narrowed. Note that the enhancement factor could be huge in real physical systems: assuming $\sigma \sim 1$, $n(\omega_{RES}) \sim 1$, and $v_G/c \approx 10^{-7}$ (as observed in a recent ultra slow-light experiment [15]), we get the enhancement factor of order *$10^7$*!

In order to confirm the validity of our model from Eq.(3), we perform a series of numerical experiments on an exemplary microcavity system. That is, we perform finite-difference-time-domain (FDTD) simulations [16], which solve Maxwell's equations exactly (with no approximation apart for the discretization), including the dispersion, with perfectly-matched-layers (PML) boundary conditions. The cavity we focus our attention on is a so-called monorail photonic crystal microcavity, shown in top plot of Figure 1. It consists of a periodically corrugated waveguide; the cavity is introduced by introducing a defect into the periodicity. The signal is sent down the waveguide on the left (which serves as the input channel); it couples through tunneling into the cavity, from where it decays into the radiation modes, and also into the waveguide on the left, and the waveguide on the right (which serves as the output channel). A monorail cavity of this class has actually already been experimentally implemented in an *Si/SiO$_2$* system [17], with resonant wavelength of *1.56μm*, *Q=265*, peak transmission of *82%*, and modal volume *0.055μm$^3$*. FDTD numerical simulations [17] of that system reproduced all experimental features very faithfully, and even quantitatively up to a discrepancy of only a few percent. Since in the current work we are not interested in a particular physical system but rather in studying the underlying physical phenomena, we can reduce our numerical requirements immensely by studying a 2D (instead of 3D) system. 3D FDTD simulations would be prohibitively time-consuming in the regime of large life-times that we are interested in, while we do not expect the physics of our particular 2D model to be any different than its 3D counterpart [18]. In all our simulations, the numerical resolution is *40pts/a*.

As our first numerical simulation, we "measure" the transmission through our system, in the case when *n=3.464* in the "central" region. The modal profile of the resonant mode is shown in the bottom panel of Figure 1. The band-gap extends from $\omega \approx 0.2(2\pi c/a)$ to $\omega \approx 0.35(2\pi c/a)$. As shown by the blue curve in Figure 2, the resonant frequency of the cavity occurs at $\omega_{RES}=0.25443(2\pi c/a)$, $Q_{TRANS}=308$, and on-resonant transmission is $T_{PEAK} \equiv P_{OUT}(\omega_{RES})/P_{IN}(\omega_{RES})=0.7597$. Since this model does not include absorption, the transmission is limited to the value below *100%* because of the coupling to the radiation modes. From these simulations, we conclude that $Q_{RAD} = Q_{TRANS} / \left(1 - \sqrt{T_{PEAK}}\right) = 2399$.

Next, we introduce material dispersion into the "central" region of our system: its $n(\omega)$ is shown in Figure 3 [19]. The system is designed so that $Re\{n(\omega_{RES})\}= n_{Si}(\lambda=1.5\mu m)=3.464$, $Im\{n(\omega_{RES})\}=0$, while $v_G(\omega_{RES})/c=0.0453$. (This particular value of $v_G$ makes comparison with a comparable non-dispersive cavity easier, as we will see later). Over the frequency range of interest, dispersion is almost linear (so $v_G$ is nearly constant), while absorption is very small. When we calculate the transmission through this system, we obtain the solid green curve in Figure 2, which has: $\omega_{RES}=0.25446(2\pi c/a)$, $T_{PEAK}=0.7613$, and $Q_{TRANS}=1106$. This value of $Q_{TRANS}$ is consistent with the one obtained by plugging $v_G(\omega_{RES})/c=0.0453$, $Q_{TRANS}=308$ (from the blue curve above), and $\sigma=0.425$, (which is obtained with a numerical computation that is independent of the other computations) into the perturbation theory result given by Eq.(4) [20]. Despite the fact that light now spends much more time in the cavity (thereby having more time to couple to the radiation modes), the peak transmission is not affected. To appreciate the significance of this fact, consider an alternative (very commonly used) way of increasing lifetime: instead of adding material dispersion, we add one more period of holes to the sides of the cavity (so there are 4 holes on each side). The transmission is shown as the solid red curve in Figure 2: $Q_{TRANS}=1079$ (this is very similar to $Q_{TRANS}$ in the case of the green curve because of the particular $v_G$ value chosen for the green curve), but $T_{PEAK}=0.2918$, which is *2.6* times lower than for the solid green curve. In the temporal domain (not shown), both the green, and red solid curves are exponentially decaying with the same rate, but in the case of the green curve, ≈*2.6* times more energy is transmitted to the output than in the case of the red curve. Intuitively, one can also understand this peculiar influence of the material dispersion as follows: from the point of view of material dispersion, both the radiation modes, and waveguide modes look the same. Therefore dispersion weakens (slows down) the coupling to each of these modes by the same amount, thus making $Q_{TRANS}$ larger, while leaving $T_{PEAK}$ unchanged. In contrast, adding an additional hole to each side of the cavity lowers only $\Gamma_{IO}$, thus decreasing $T_{PEAK}$. Before proceeding, we perform one final check on this picture by lowering $v_G(\omega_{RES})/c$ in the central region to *0.0150*. The result is shown as the solid black curve in Figure 2: although $T_{PEAK}$ is nearly the same as the solid green and blue curves, $Q_{TRANS}=2665$, which is again a value consistent with the perturbation theory result [20], despite the fact that at such low group velocity we are stretching the limits of our numerical resolution. Note however that because of the extreme resonance width narrowing at such a low group velocity, only frequencies $\omega$ that are *very* close to $\omega_{RES}$ play important role for the system. But, it is precisely for these frequencies that our perturbative model from Eq.(3) is most justified, since all our expansions become better and better approximations precisely in that limit. For example, neglecting second order dispersion in Eq.(2) causes even smaller errors than in the case when group velocity is larger.

Considerations of the previous paragraph were for nearly absorption-free systems. Now, we proceed to study the effects of increased material absorption. To do this, we take the exact systems presented by the solid blue, red, and green curves in Figure 2, and add the same amount of absorption $Im\{n\}=0.0077$ to each of them. The resulting transmissions are shown by the dashed curves in Figure 2. Consistent with our model, the $T_{PEAK}$s in the blue and green case decreased by the same factor ($T_{PEAK}\approx0.276$ now); both of these curves also have significantly lower $Q_{TRANS}$s now ($Q_{TRANS}$=*187*, and *669* respectively), but as expected by Eq.(4), the ratio of their $Q_{TRANS}$s did not change. Finally, from $T_{PEAK}$, $Q_{RAD}$ and $Q_{TRANS}$ of the blue curve, we obtain $Q_{ABS}$=*476*, which is quantitatively consistent with the observed $T_{PEAK}$=*0.0273* for the dashed red curve, and with the perturbation theory prediction $Q_{ABS}=n/(2Im\{n\}\sigma)$ [20]. Since the light has much more time to interact with the absorptive material in the case of the dashed red curve ($Q_{TRANS}$=*327*), the resulting factor of decrease in $T_{PEAK}$ is much larger than in the case of the dashed blue curve. In contrast, our dispersive cavity (even now, in presence of absorption) has 10 times larger transmission than the dashed red curve, while actually having a larger lifetime (by a factor of 2). Another way to understand this somewhat counter-intuitive result is to note that at $\omega_{RES}$, the systems corresponding to the blue and green curves look exactly the same, meaning that transmission at $\omega_{RES}$ has to be exactly the same; the rest of our results follow from this simple constraint.

Before concluding, a few words are in order to discuss various possible physical implementations of the scheme we propose. The ultra slow-light experiment from Ref. 15, is an obvious option. There are other potentially promising systems to create slow-light media, including: polaritonic media, metal-dielectric surfaces supporting surface-plasmons, and USL in solids [23]. Polaritons and surface-plasmons tend to be very lossy in the regimes of high dispersion; to compensate for the loss, they could be combined with gain media, or they could be cooled to very low temperatures in which case losses drop dramatically. In USL losses are not a problem, and it has a further interesting characteristic that its dispersion can be externally controlled with changing the amplitude of an external coupling field [15]; this could potentially provide a micro-cavity whose Q could be dynamically changed by many orders of magnitude. A typical *n(ω)* in USL media has a very similar shape to the one we show in Figure 3; introducing USL media into a cavity would therefore produce very similar results to the ones we show in Figure 2. Because of the importance of their applications, very significant efforts have been devoted to designing high-Q microcavities [8,21,22]. Most of these designs are compatible with using highly-dispersive materials; gaseous USL is suitable for use with a photonic crystal microcavity from Ref. 21, while solid-state based (in *Pr* doped $Y_2SiO_5$) USL [23] could be used with most existing microresonator designs. For example, by naively combining solid-state USL from Ref.23 ($v_G/c=1.5*10^{-7}$) with the microcavity design from Ref.8 ($Q>10^8$), it may be possible to achieve $Q>10^{15}$. (Clearly, to achieve such long lifetimes, many various technical hurdles would have to be overcome, including: time-dependent stray fields, and temperature fluctuations). Alternatively, recently demonstrated surface-plasmon photonic crystals [24], or photonic crystals incorporating polaritons [25], could also be used. In any case, the new physical phenomena presented here should have relevance to a variety of important applications.

In conclusion, we have demonstrated how adding highly dispersive materials to microresonators could be used to increase their lifetimes by many orders of magnitude, while preserving their peak transmissions, even in the presence of radiation, and absorption losses. Microresonators of such unprecedently narrow line-widths, and tiny modal volumes, might enable many important applications in fields as diverse as: cavity QED, nonlinear optics, and perhaps even integrable atomic clocks. The underlying physical mechanism applies for any cavity geometry, but we expect it to be most beneficial precisely in the case of microresonators since they are typically most limited by radiation and absorption losses.

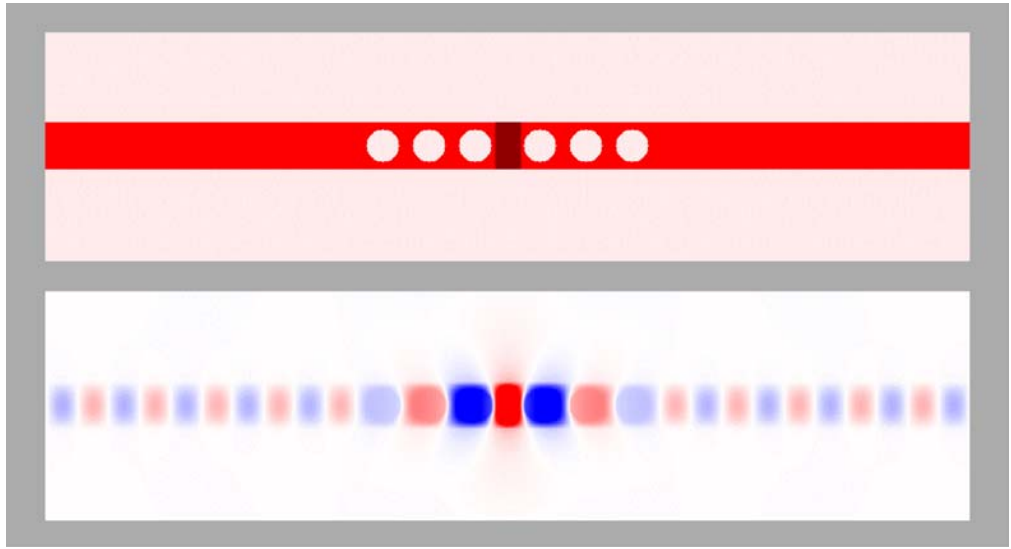

**Figure 1:** The photonic crystal microcavity system used in all our numerical experiments. The top panel is a sketch of the system (red denotes *n=3.464*, the rest is *n=1*). If we denote the thickness of the monorail with *a*, then the distance between successive holes is also *a*, except for the defect in the periodicity which presents the cavity, where the distance between the holes is increased to *1.4a*. The radius of each hole is *0.35a*. In various numerical simulations in this manuscript, we change only the properties of the "central", shaded region of this structure; the thickness of this region is the same as the monorail, while its width is *0.6a*. The bottom panel shows the magnetic field of the confined mode, which is perpendicular to the plane everywhere, while the electric field lies in the plane.

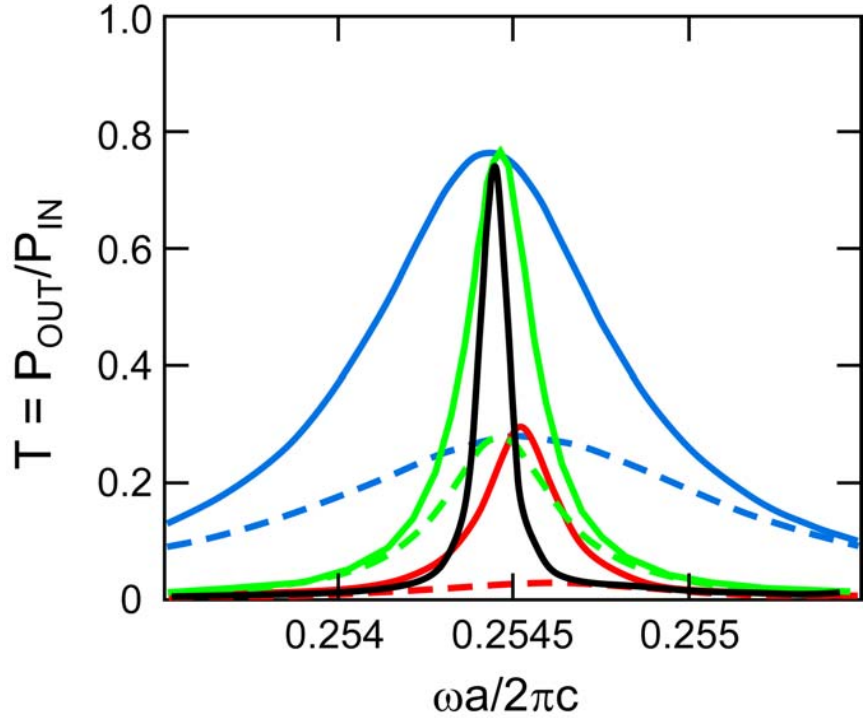

**Figure 2:** Transmission curves for all the microcavities (sketched in Figure 1) that were simulated. Solid blue and red lines denote transmission through cavities made from non-dispersive material: blue is for the case with three holes on each side of the cavity, while red is for the case of four holes on each side. Solid green and black curves denote transmission through cavities in which dispersive material was included in the shaded "central" region of Figure 1; in each case, there were 3 holes on each side of the cavity. Green curve is for the cavity whose $n(\omega)$ is shown in Figure 3; it has $v_G(\omega_{RES})/c=0.0453$. Black curve is for the cavity that has $v_G(\omega_{RES})/c=0.0150$. Dashed curves are for the cavities that are exactly the same as their corresponding-color solid curve counterparts, but now also including absorption: $Im\{n\}=0.0077$.

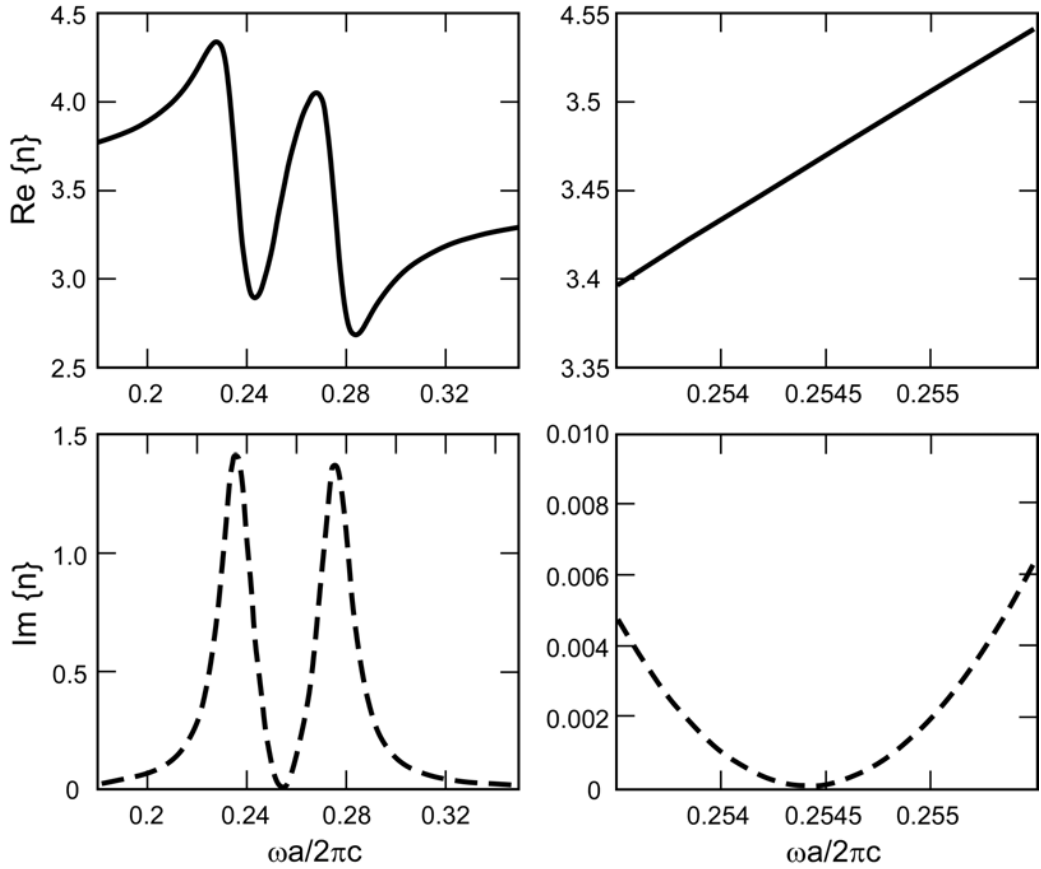

**Figure 3:** Example of an *n(ω)* used in our simulations; both imaginary and real parts are shown. The column on the left shows *n(ω)* over a broad frequency range. The column on the right shows the same *n(ω)*, but over the frequency regime relevant for the microcavities studied. In the case shown here, $v_G/c \approx 0.0453$ in the region of interest.